\pdfoutput=1
\documentclass[12pt]{article}
\usepackage{amsmath,amssymb,amsfonts,here}
\usepackage{tikz}
\begin{document}
\newtheorem{theorem}{Theorem}[section]
\newtheorem{corollary}[theorem]{Corollary}
\newtheorem{lemma}[theorem]{Lemma}
\newtheorem{remark}[theorem]{Remark}
\newtheorem{example}[theorem]{Example}
\newtheorem{proposition}[theorem]{Proposition}
\newtheorem{definition}[theorem]{Definition}
\newtheorem{assumption}[theorem]{Assumption}
\def\emptyset{\varnothing}
\def\setminus{\smallsetminus}
\def\id{{\mathrm{id}}}
\def\G{{\mathcal{G}}}
\def\E{{\mathcal{E}}}
\def\H{{\mathcal{H}}}
\def\C{{\mathbb{C}}}
\def\N{{\mathbb{N}}}
\def\Q{{\mathbb{Q}}}
\def\R{{\mathbb{R}}}
\def\Z{{\mathbb{Z}}}
\def\Path{{\mathrm{Path}}}
\def\Str{{\mathrm{Str}}}
\def\st{{\mathrm{st}}}
\def\tr{{\mathrm{tr}}}
\def\opp{{\mathrm{opp}}}
\def\a{{\alpha}}
\def\be{{\beta}}
\def\de{{\delta}}
\def\e{{\varepsilon}}
\def\si{{\sigma}}
\def\la{{\lambda}}
\def\th{{\theta}}
\def\lan{{\langle}}
\def\ran{{\rangle}}
\def\isom{{\cong}}
\newcommand{\Hom}{\mathop{\mathrm{Hom}}\nolimits}
\newcommand{\End}{\mathop{\mathrm{End}}\nolimits}
\def\qed{{\unskip\nobreak\hfil\penalty50
\hskip2em\hbox{}\nobreak\hfil$\square$
\parfillskip=0pt \finalhyphendemerits=0\par}\medskip}
\def\proof{\trivlist \item[\hskip \labelsep{\bf Proof.\ }]}
\def\endproof{\null\hfill\qed\endtrivlist\noindent}

%%%%%%%%%%%%%%%%%%%%%%%%%%%%%%
\title{Quantum $6j$-symbols arising from a subfactor\\
of finite index and finite depth and\\
the zipper condition for $4$-tensors\\
in two-dimensional topological order}
\author{
{\sc Yasuyuki Kawahigashi}\\
{\small iTHEMS Research Group, RIKEN}\\
{\small 2-1 Hirosawa, Wako, Saitama 351-0198,Japan}\\
{\small e-mail: {\tt yasuyuki.kawahigashi@riken.jp}}
\\[0,40cm]
{\small Graduate School of Mathematical Sciences}\\
{\small The University of Tokyo, Komaba, Tokyo, 153-8914, Japan}
\\[0,05cm]
{\small and}
{\small Kavli IPMU (WPI), the University of Tokyo}\\
{\small 5--1--5 Kashiwanoha, Kashiwa, 277-8583, Japan}}
\maketitle{}

\begin{abstract}
We study 3-tensors satisfying the ``zipper condition,'' 
which play an important role in tensor network approaches 
to two-dimensional topological order.  In our previous 
work, we identified the 4-tensors appearing in these 
approaches with bi-unitary connections in the Jones theory 
of subfactors in operator algebras.  These bi-unitary 
connections give rise to subfactors of finite index.  We 
prove that, when the associated subfactor has finite depth, 
the 3-tensors satisfying the zipper condition correspond 
precisely to the quantum $6j$-symbols, also known as 
$F$-symbols, of the unitary multi-fusion category arising 
from the subfactor.  This realization of quantum $6j$-symbols 
differs from their realization through a flat bi-unitary 
connection arising from the higher relative commutants 
of a subfactor.
\end{abstract}

\section{Introduction}

Two-dimensional topological order has attracted 
considerable interest in both physics and mathematics. 
From a mathematical perspective, the unitary braided 
fusion categories that arise in this context are closely 
connected to a broad range of topics, including quantum 
groups, quantum invariants in low-dimensional topology, 
integrable lattice models, tensor networks, conformal 
field theory, vertex operator algebras, and operator 
algebras. In this paper, we investigate the relationship 
between tensor networks and subfactor theory within the 
framework of operator algebras.

Tensor categories \cite{EGNO} provide a mathematical 
framework for quantum symmetries that extends the classical 
framework based on groups. We are particularly interested 
in unitary fusion categories, which serve as counterparts 
of finite groups in this broader theory of symmetry. 
Unitary modular tensor categories form a particularly 
important class of unitary fusion categories. The celebrated 
Drinfeld center construction associates a unitary modular 
tensor category with any unitary fusion category.

Tensor network approaches to two-dimensional topological 
order based on fusion categories have become an active area 
of research in mathematical physics. In particular, the 
authors of \cite{BMWSHV} study unitary fusion categories 
and their Drinfeld centers within a tensor network framework 
formulated in terms of 3-tensors and 4-tensors.

The subfactor theory of Jones provided one of the earliest 
mathematical frameworks in which quantum symmetries appeared. 
A subfactor is an inclusion $N\subset M$ of certain 
infinite-dimensional operator algebras on a Hilbert space
called factors \cite{EK1}. Under the finiteness assumptions 
of finite index and finite depth, such an inclusion gives 
rise to unitary fusion categories of $N$-$N$ and $M$-$M$ 
bimodules, which encode the symmetry of $N\subset M$. 
Jones's discovery of the underlying structure \cite{J1} 
subsequently led to the celebrated Jones polynomial, an 
invariant of links \cite{J2}.

Subfactor theory was later found to have deep connections 
with many important topics in the operator algebraic 
approach to mathematical physics, including quantum 
statistical mechanics and quantum field theory \cite{EK1}, 
\cite{EK2}, \cite{KL}, \cite{L1}, \cite{L2}. More recent 
developments can be found in \cite{Jo}, \cite{JL}, \cite{Ko}, 
\cite{Og}. In particular, exotic fusion categories, such 
as the Haagerup category \cite{AH}, which appears to lie 
beyond previously known constructions, have recently 
attracted substantial interest in connection with conformal 
field theory \cite{I}, \cite{Hu}, \cite{R}.

Ocneanu introduced bi-unitary connections \cite{O1}, 
\cite{O2} as a tool for classifying subfactors in terms of 
finite-dimensional data \cite{P2}. The notion was subsequently 
generalized in \cite{Sc} to the study of commuting squares 
\cite{P1}, and its interpretation as an object in a unitary 
fusion category was established in \cite{AH}. See also 
\cite{EK1}, \cite{K1}. More recently, bi-unitary connections 
have been investigated from several perspectives in \cite{BC}, 
\cite{CEM}, \cite{DGGJ}.

A direct relationship between the 4-tensors of \cite{BMWSHV} 
and bi-unitary connections was pointed out in \cite{K2}. An 
operator algebraic interpretation of matrix product operators 
\cite{BMWSHV}, \cite{H}, \cite{LFHSV}, \cite{MRGSCP}, 
\cite{RGMP} was subsequently developed in \cite{K3}. A 
review of these connections is given in \cite{K4}.

More recently, the zipper condition, an important condition 
expressing a morphism property of the 4-tensors in \cite{BMWSHV}, 
was identified in \cite{K4} with the notion of flat fields 
of strings \cite[Section 11.4]{EK1} in subfactor theory. 
Although the zipper condition can naturally be seen 
as a certain intertwining relation in a fusion category, its precise 
formulation in subfactor-theoretic terms has remained unclear. 
The main result, Theorem \ref{main}, 
of this paper establishes a direct and 
explicit relationship between these two formulations
by identifying the coefficients
appearing in the zipper condition and the quantum $6j$-symbols
in the finite depth case. 
To this end, we use results 
from our previous work \cite{K5}, which was originally 
undertaken for a different purpose. Specifically, we 
characterize the 3-tensors satisfying the zipper condition 
in terms of a 4-tensor arising from certain quantum 
$6j$-symbols, also known as $F$-symbols.

\section{Preliminaries on bi-unitary connections}

We set preliminaries on a bi-unitary connection, which gives
a characterization of a finite dimensional non-degenerate
commuting square.  See \cite[Section 11.2]{EK1} for a
general treatment in subfactor theory and
\cite{K6} for the current framework.

Let $L\subset N \subset M \subset P\subset Q$ be
inclusions of type II$_1$ factors with $1 < [Q:L] < \infty$
and $L\subset Q$ having a finite depth.  Here some of
the inclusions can be identities.  (When we construct
examples, these factors are hyperfinite, but we do not
need this assumption below.)  We assume that all the global indices
of non-trivial subfactors arising from these inclusions are
equal.  (Recall that the global index of $N\subset M$ is
defined to be $\sum_{{}_N X_N} (\dim_N X)^2$, where ${}_N X_N$
runs through all irreducible $N$-$N$ bimodules generated
by ${}_N M_N$.)  These equalities imply that the $L$-$L$ bimodules
arising from $L\subset N$ and those from $L\subset Q$ are the
same as long as $L\neq N$, for example.  
Consider bimodules with finite index
arising from these subfactors.  Since we have at most
five factors, we have at most 25 kinds of bimodules.  Because
of the finite depth assumption, we have a unitary
multi-fusion category of these bimodules.
(Each has a finite index automatically.)
We can apply the following results in this paper also
to a multi-fusion category of bimodules of more than 25 kinds,
but it is enough to consider a multi-fusion full subcategory
of at most 25 kinds of bimodules, so our setting is
sufficiently general.

For $R_1, R_2\in\{L,N,M,P,Q\}$ and an $R_1$-$R_2$ bimodule
${}_{R_1}X_{R_2}$ in the 
multi-fusion category, we simply write $X$ and set
$d(X)=(\dim {}_{R_1}X \dim{}X_{R_2})^{1/2}$.

Fix a (possibly reducible) ${}_P D_Q$ in the multi-fusion category.
For an irreducible bimodule ${}_N A_M$ in this multi-fusion category
and irreducible bimodules ${}_M (X_1)_P$,
${}_M (X_2)_Q$,
${}_N (X_3)_P$, and
${}_N (X_4)_Q$ in this multi-fusion category, 
we define a bi-unitary connection $W_A$
as in Fig.~\ref{connA} where
$S_1\in\Hom(A\otimes X_1, X_3)$,
$S_4\in\Hom(X_1\otimes D, X_2)$,
$S_5\in\Hom(X_3\otimes D, X_4)$, and
$S_7\in\Hom(A\otimes X_2, X_4)$ are coisometric intertwiners.
(We choose $S_k$ so that we have an orthonormal basis of
each Hom space here and later.)

Here Hom represents the space of appropriate bimodule
intertwiners.  For example, $\Hom(A\otimes X_1, X_3)$
actually means $\Hom({}_N A\otimes_M (X_1)_P, {}_N (X_3)_P)$.
We also use a convention that we simply write $S_1$ instead
of $S_1\otimes \id_D$ for an intertwiner from $A\otimes X_1\otimes D$
to $X_3\otimes D$.  This convention applies to all similar
cases in this paper.  The above definition indeed gives
a bi-unitary connection because the values of $W_A$ are
special values of the quantum $6j$-symbols as considered
in \cite[Section 12.4]{EK1}.  We considered six bimodules
over only two factors in
\cite[Section 12.4]{EK1}, but the same arguments 
apply to our case here.

Similarly, for irreducible bimodules ${}_L B_N$ and ${}_L C_M$
in this multi-fusion category,
we define bi-unitary connections $W_B$ and $W_C$
as in Fig.~\ref{connBC} where
${}_L (X_5)_P$ and
${}_L (X_6)_Q$ are irreducible bimodules in this multi-fusion
category and
$S_2\in\Hom(C\otimes X_1, X_5)$,
$S_3\in\Hom(B\otimes X_3, X_5)$,
$S_6\in\Hom(X_5\otimes D, X_6)$,
$S_8\in\Hom(B\otimes X_4, X_6)$, and
$S_9\in\Hom(C\otimes X_2, X_6)$,
are coisometric intertwiners.  We choose $D$ so that all
horizontal graphs for $W_A, W_B, W_C$ are connected.
This is possible because we allow $D$ to be reducible.

We fix coisometric intertwiners corresponding to 
common horizontal
edges of $W_A, W_B, W_C$, but we can change choices
of coisometric intertwiners corresponding to 
vertical edges of $W_A$, $W_B$ or $W_C$, and then
the corresponding bi-unitary connection changes within its
equivalence class.  See \cite[Theorem 3]{AH} and
the Remark after that for this type of gauge choices
and equivalence for bi-unitary connections.

We can also use $(B\otimes A)\oplus C$ for the vertical
tensor product and then obtain a bi-unitary connection
$W_{(B\otimes A)\oplus C}$ in a similar way.
(The bimodule $(B\otimes A)\oplus C$ is reducible,
but this does not matter.)  Then this
bi-unitary connection $W_{(B\otimes A)\oplus C}$ corresponds
to $(W_B\otimes W_A)\oplus W_C$, where the
tensor product and the direct sum in
$(W_B\otimes W_A)\oplus W_C$ are those for
bi-unitary connections as in \cite[Section 3]{AH}.

\begin{figure}[H]
\begin{center}
\begin{tikzpicture}
\draw [thick, ->] (2.5,2.5)--(4,2.5);
\draw [thick, ->] (1,2)--(1,1.5);
\draw [thick, ->] (5,2)--(5,1.5);
\draw [thick, ->] (2,1)--(4.5,1);
\draw (3.25,2.5)node[above]{$S_4$};
\draw (1,1.75)node[left]{$S_1$};
\draw (5,1.75)node[right]{$S_7$};
\draw (3.25,1)node[below]{$S_5$};
\draw (3,1.75)node{$W_A$};
\draw (1,2.5)node{$A\otimes X_1\otimes D$};
\draw (5,2.5)node{$A\otimes X_2$};
\draw (1,1)node{$X_3 \otimes D$};
\draw (5,1)node{$X_4$};
\end{tikzpicture}
\caption{The bi-unitary connection arising from $A$}
\label{connA}
\end{center}
\end{figure}

\begin{figure}[H]
\begin{center}
\begin{tikzpicture}
\draw [thick, ->] (2.5,2.5)--(4,2.5);
\draw [thick, ->] (10.5,2.5)--(12,2.5);
\draw [thick, ->] (1,2)--(1,1.5);
\draw [thick, ->] (5,2)--(5,1.5);
\draw [thick, ->] (9,2)--(9,1.5);
\draw [thick, ->] (13,2)--(13,1.5);
\draw [thick, ->] (2,1)--(4.5,1);
\draw [thick, ->] (10,1)--(12.5,1);
\draw (3.25,2.5)node[above]{$S_5$};
\draw (11.25,2.5)node[above]{$S_4$};
\draw (1,1.75)node[left]{$S_3$};
\draw (5,1.75)node[right]{$S_8$};
\draw (9,1.75)node[left]{$S_2$};
\draw (13,1.75)node[right]{$S_9$};
\draw (3.25,1)node[below]{$S_6$};
\draw (11.25,1)node[below]{$S_6$};
\draw (3,1.75)node{$W_B$};
\draw (11,1.75)node{$W_C$};
\draw (1,2.5)node{$B\otimes X_3\otimes D$};
\draw (5,2.5)node{$B\otimes X_4$};
\draw (9,2.5)node{$C\otimes X_1\otimes D$};
\draw (13,2.5)node{$C\otimes X_2$};
\draw (1,1)node{$X_5 \otimes D$};
\draw (5,1)node{$X_6$};
\draw (9,1)node{$X_5\otimes D$};
\draw (13,1)node{$X_6$};
\end{tikzpicture}
\caption{The bi-unitary connections arising from $B$ and $C$}
\label{connBC}
\end{center}
\end{figure}

We recall that a non-degenerate
finite-dimensional commuting square that produces a 
subfactor of finite index and finite depth needs to be
of the  above form by \cite[Theorem 3.4]{K5}.
That is, in our main Theorem \ref{main} below, we consider
a bi-unitary connection of this type, but it is
equivalent to having a commuting square producing
a subfactor of finite index and finite depth.

For a diagram in Fig.~\ref{connA}, we simply drop $A$ and $D$
when we write bimodules and have a bi-unitary connection 
$W_A$.  This is a standard convention in writing a bi-unitary
connection.  We further
define a $4$-tensor $a_A$ using it as in Fig.~\ref{tensorA}
as in \cite[Fig.~11]{K2}.  (We drop labels for vertices of
the connection $W_A$, but this information is kept for
labels of $S_1,S_4,S_5,S_7$.)
We define $4$-tensors $a_B$ and $a_C$ similarly.

\begin{figure}[H]
\begin{center}
\begin{tikzpicture}
\draw [thick] (2.5,1)--(2.5,1.2);
\draw [thick] (2.5,2)--(2.5,1.8);
\draw [thick] (2,1.5)--(2.2,1.5);
\draw [thick] (3,1.5)--(2.8,1.5);
\draw (2.5,1.5) circle (0.3);
\draw (2.5,1.5)node{$a_A$};
\draw (2,1.5)node[left]{$S_1$};
\draw (3,1.5)node[right]{$S_7$};
\draw (2.5,1)node[below]{$S_5$};
\draw (2.5,2)node[above]{$S_4$};
\draw (5.5,1.6)node{$\displaystyle=
\sqrt[4]{\frac{d(X_1))d(X_4)}{d(X_2))d(X_3)}}$};
\draw [thick, ->] (8.5,2.5)--(9.5,2.5);
\draw [thick, ->] (8,2)--(8,1.5);
\draw [thick, ->] (10,2)--(10,1.5);
\draw [thick, ->] (8.5,1)--(9.5,1);
\draw (9,2.5)node[above]{$S_4$};
\draw (8,1.75)node[left]{$S_1$};
\draw (10,1.75)node[right]{$S_7$};
\draw (9,1)node[below]{$S_5$};
\draw (9,1.75)node{$W_A$};
\draw (8,2.5)node{$X_1$};
\draw (10,2.5)node{$X_2$};
\draw (8,1)node{$X_3$};
\draw (10,1)node{$X_4$};
\end{tikzpicture}
\caption{The $4$-tensor $a_A$ and the bi-unitary connection $W_A$}
\label{tensorA}
\end{center}
\end{figure}

\section{Main results}

Let $T\in\Hom(B\otimes A, C)$ be a coisometric intertwiner.
We define quantum $6j$-symbols $W_T$ and $W'_T$ as in
Fig.~\ref{q6T}.

\begin{figure}[H]
\begin{center}
\begin{tikzpicture}
\draw [thick, ->] (2.5,2.5)--(4,2.5);
\draw [thick, ->] (10.5,2.5)--(12,2.5);
\draw [thick, ->] (1,2)--(1,1.5);
\draw [thick, ->] (5,2)--(5,1.5);
\draw [thick, ->] (9,2)--(9,1.5);
\draw [thick, ->] (13,2)--(13,1.5);
\draw [thick, ->] (2,1)--(4.5,1);
\draw [thick, ->] (10,1)--(12.5,1);
\draw (3.25,2.5)node[above]{$S_1$};
\draw (11.25,2.5)node[above]{$S_7$};
\draw (1,1.75)node[left]{$T$};
\draw (5,1.75)node[right]{$S_3$};
\draw (9,1.75)node[left]{$T$};
\draw (13,1.75)node[right]{$S_8$};
\draw (3.25,1)node[below]{$S_2$};
\draw (11.25,1)node[below]{$S_9$};
\draw (3,1.75)node{$W_T$};
\draw (11,1.75)node{$W'_T$};
\draw (1,2.5)node{$B\otimes A\otimes X_1$};
\draw (5,2.5)node{$B\otimes X_3$};
\draw (9,2.5)node{$B\otimes A\otimes X_2$};
\draw (13,2.5)node{$B\otimes X_4$};
\draw (1,1)node{$C \otimes X_1$};
\draw (5,1)node{$X_5$};
\draw (9,1)node{$C\otimes X_2$};
\draw (13,1)node{$X_6$};
\end{tikzpicture}
\caption{Quantum $6j$-symbols arising from $T$}
\label{q6T}
\end{center}
\end{figure}

We define a $3$-tensor $F_T$ using $W_T$ as in Fig.~\ref{threeF}.

\begin{figure}[H]
\begin{center}
\begin{tikzpicture}[scale=1.2]
\draw [thick] (1.7,1)--(2,1);
\draw [thick] (1.7,2)--(2,2);
\draw [thick] (1.3,1.5)--(1,1.5);
\draw [thick] (1.3,0.7)--(1.7,0.7);
\draw [thick] (1.3,2.3)--(1.7,2.3);
\draw [thick] (1.3,0.7)--(1.3,2.3);
\draw [thick] (1.7,0.7)--(1.7,2.3);
\draw (1.5,1.5)node{$F_T$};
\draw (1,1.5)node[left]{$S_2$};
\draw (2,2)node[right]{$S_1$};
\draw (2,1)node[right]{$S_3$};
\draw (3.5,1.5)node{$=\displaystyle\frac{d(X_5)}{d(X_1)}$};
\draw [thick, ->] (7,2)--(8.5,2);
\draw [thick, ->] (5.5,1.75)--(5.5,1.25);
\draw [thick, ->] (9.5,1.75)--(9.5,1.25);
\draw [thick, ->] (6.5,1)--(9,1);
\draw (7.75,2)node[above]{$S_1$};
\draw (5.5,1.5)node[left]{$T$};
\draw (9.5,1.5)node[right]{$S_3$};
\draw (7.75,1)node[below]{$S_2$};
\draw (7.5,1.5)node{$W_T$};
\draw (5.5,2)node{$B\otimes A\otimes X_1$};
\draw (9.5,2)node{$B\otimes X_3$};
\draw (5.5,1)node{$C \otimes X_1$};
\draw (9.5,1)node{$X_5$};
\end{tikzpicture}
\caption{A $3$-tensor $F_T$}
\label{threeF}
\end{center}
\end{figure}

\begin{figure}[H]
\begin{center}
\begin{tikzpicture}[scale=1.2]
\draw [thick] (1.7,1)--(2,1);
\draw [thick] (1.7,2)--(2,2);
\draw [thick] (1.3,1.5)--(1,1.5);
\draw [thick] (1.3,0.7)--(1.7,0.7);
\draw [thick] (1.3,2.3)--(1.7,2.3);
\draw [thick] (1.3,0.7)--(1.3,2.3);
\draw [thick] (1.7,0.7)--(1.7,2.3);
\draw (1.5,1.5)node{$F'_T$};
\draw (1,1.5)node[left]{$S_9$};
\draw (2,2)node[right]{$S_7$};
\draw (2,1)node[right]{$S_8$};
\draw (3.5,1.5)node{$=\displaystyle\frac{d(X_6)}{d(X_2)}$};
\draw [thick, ->] (7,2)--(8.5,2);
\draw [thick, ->] (5.5,1.75)--(5.5,1.25);
\draw [thick, ->] (9.5,1.75)--(9.5,1.25);
\draw [thick, ->] (6.5,1)--(9,1);
\draw (7.75,2)node[above]{$S_7$};
\draw (5.5,1.5)node[left]{$T$};
\draw (9.5,1.5)node[right]{$S_8$};
\draw (7.75,1)node[below]{$S_9$};
\draw (7.5,1.5)node{$W'_T$};
\draw (5.5,2)node{$B\otimes A\otimes X_2$};
\draw (9.5,2)node{$B\otimes X_4$};
\draw (5.5,1)node{$C \otimes X_2$};
\draw (9.5,1)node{$X_6$};
\end{tikzpicture}
\caption{A $3$-tensor $F'_T$}
\label{threeFp}
\end{center}
\end{figure}
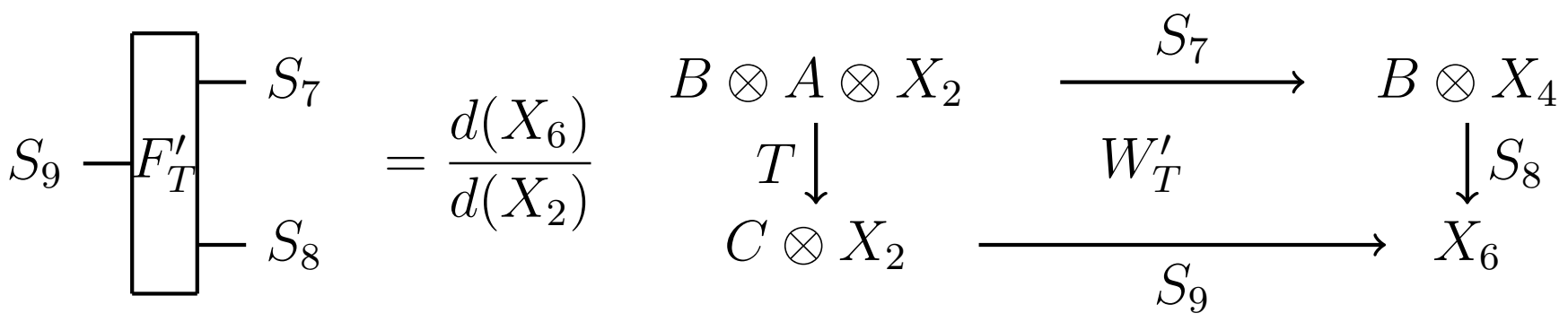

We also define a $3$-tensor $F'_T$ using $W'_T$ similarly
as in Fig.~\ref{threeFp}.

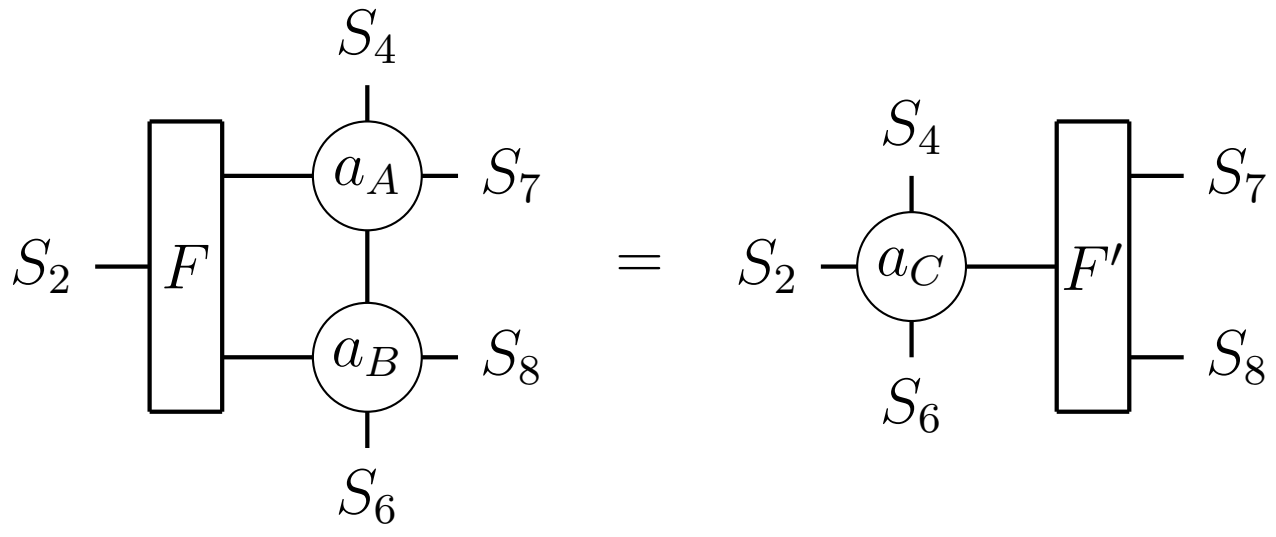
\begin{figure}[H]
\begin{center}
\begin{tikzpicture}[scale=1.2]
\draw [thick] (2.5,1.5)--(2.5,1.7);
\draw [thick] (2.5,2.7)--(2.5,2.3);
\draw [thick] (2.5,3.5)--(2.5,3.3);
\draw [thick] (1.7,2)--(2.2,2);
\draw [thick] (3,2)--(2.8,2);
\draw [thick] (1.7,3)--(2.2,3);
\draw [thick] (3,3)--(2.8,3);
\draw (2.5,3) circle (0.3);
\draw (2.5,2) circle (0.3);
\draw (2.5,3)node{$a_A$};
\draw (2.5,2)node{$a_B$};
\draw [thick] (1.3,1.7)--(1.7,1.7);
\draw [thick] (1.3,3.3)--(1.7,3.3);
\draw [thick] (1.3,1.7)--(1.3,3.3);
\draw [thick] (1.7,1.7)--(1.7,3.3);
\draw [thick] (1.3,2.5)--(1,2.5);
\draw (1.5,2.5)node{$F$};
\draw (1,2.5)node[left]{$S_2$};
\draw (2.5,3.5)node[above]{$S_4$};
\draw (3,3)node[right]{$S_7$};
\draw (3,2)node[right]{$S_8$};
\draw (2.5,1.5)node[below]{$S_6$};
\draw (4,2.5)node{$=$};
\draw [thick] (5.5,2)--(5.5,2.2);
\draw [thick] (5.5,3)--(5.5,2.8);
\draw [thick] (5,2.5)--(5.2,2.5);
\draw [thick] (6.3,2.5)--(5.8,2.5);
\draw (5.5,2.5) circle (0.3);
\draw (5.5,2.5)node{$a_C$};
\draw [thick] (6.3,1.7)--(6.7,1.7);
\draw [thick] (6.3,3.3)--(6.7,3.3);
\draw [thick] (6.3,1.7)--(6.3,3.3);
\draw [thick] (6.7,1.7)--(6.7,3.3);
\draw [thick] (6.7,2)--(7,2);
\draw [thick] (6.7,3)--(7,3);
\draw (6.5,2.5)node{$F'$};
\draw (5,2.5)node[left]{$S_2$};
\draw (7,3)node[right]{$S_7$};
\draw (7,2)node[right]{$S_8$};
\draw (5.5,3)node[above]{$S_4$};
\draw (5.5,2)node[below]{$S_6$};
\end{tikzpicture}
\caption{The half-zipper condition for $F, F'$}
\label{halfzipper}
\end{center}
\end{figure}

We say that a pair $(F, F')$ of 3-tensors satisfies 
the half zipper condition for $a_A, a_B, a_C$ 
if it satisfies the equality in Fig.~\ref{halfzipper}.
This condition and the name have been introduced
in \cite[Fig. 36]{K6} and \cite[Theorem 4.1]{K6},
respectively.

\begin{theorem}\label{main}
A pair $(F, F')$ of 3-tensors satisfies the half zipper
condition for $a_A, a_B, a_C$ if and only if it arises 
from the quantum $6j$-symbols
given by a not necessarily coisometric
$T\in\Hom(B\otimes A, C)$ as 
in Fig.~\ref{threeF} and Fig.~\ref{threeFp}.
\end{theorem}

\begin{proof}
We first assume a pair $(F_T, F'_T)$ arises from 
a coisometric intertwiner $T\in\Hom(B\otimes A, C)$ as above.
(It is enough to consider only coisometric intertwiners.)
We rewrite the both sides of Fig.~\ref{halfzipper}
in terms of bi-unitary connections and quantum $6j$-symbols using
the normalization convention in Fig.~\ref{tensorA}
and Fig.~\ref{threeF}, then 
what we have to prove is that the complex number
given by the composed self-intertwiner 
$S_6 S_2 T S_4^* S_7^* S_8^*$ from $X_6$ to $X_6$ 
represented in Fig.~\ref{left} and that given by the 
self-intertwiner $S_6 S_2 S_4^* T S_7^* S_8^*$ 
represented in Fig.~\ref{right} are equal. 
This former value is equal
to that given by the self-intertwiner represented 
in Fig.~\ref{left1}, and it is equal to the latter value
because of commutativity of $S_4^*$ and $T$,
so we obtain the half zipper condition.

\begin{figure}[H]
\begin{center}
\begin{tikzpicture}
\draw [thick, ->] (3,4)--(2.5,4);
\draw [thick, ->] (7,4)--(7.5,4);
\draw [thick, ->] (6.5,2.5)--(7.5,2.5);
\draw [thick, ->] (6,1)--(8,1);
\draw [thick, ->] (5,3.5)--(5,3);
\draw [thick, ->] (9,3.5)--(9,3);
\draw [thick, ->] (5,2)--(5,1.5);
\draw [thick, ->] (9,2)--(9,1.5);
\draw [thick, ->] (1,3.5)--(4,1);
\draw (2.75,4)node[above]{$T$};
\draw (7.25,4)node[above]{$S_4$};
\draw (7,2.5)node[above]{$S_5$};
\draw (9,3.25)node[right]{$S_7$};
\draw (9,1.75)node[right]{$S_8$};
\draw (5,3.25)node[right]{$S_1$};
\draw (5,1.75)node[right]{$S_3$};
\draw (7,1)node[below]{$S_6$};
\draw (2,2.5)node[left]{$S_2$};
\draw (1,4)node{$C\otimes X_1\otimes D$};
\draw (5,4)node{$B\otimes A\otimes X_1\otimes D$};
\draw (9,4)node{$B\otimes A\otimes X_2$};
\draw (5,2.5)node{$B\otimes X_3\otimes D$};
\draw (9,2.5)node{$B\otimes X_4$};
\draw (5,1)node{$X_5 \otimes D$};
\draw (9,1)node{$X_6$};
\end{tikzpicture}
\caption{The left hand side}
\label{left}
\end{center}
\end{figure}

\begin{figure}[H]
\begin{center}
\begin{tikzpicture}
\draw [thick, ->] (7,2.5)--(7.5,2.5);
\draw [thick, ->] (6,1)--(8,1);
\draw [thick, ->] (2.5,2.5)--(3,2.5);
\draw [thick, ->] (2,1)--(4,1);
\draw [thick, ->] (1,2)--(1,1.5);
\draw [thick, ->] (9,2)--(9,1.5);
\draw (2.75,2.5)node[above]{$T^*$};
\draw (7.25,2.5)node[above]{$S_4$};
\draw (9,1.75)node[right]{$S_7$};
\draw (1,1.75)node[left]{$S_2$};
\draw (3,1)node[below]{$S_6$};
\draw (7,1)node[below]{$S_8^*$};
\draw (1,2.5)node{$C\otimes X_1\otimes D$};
\draw (5,2.5)node{$B\otimes A\otimes X_1\otimes D$};
\draw (9,2.5)node{$B\otimes A \otimes X_2$};
\draw (1,1)node{$X_5 \otimes D$};
\draw (5,1)node{$X_6$};
\draw (9,1)node{$B\otimes X_4$};
\end{tikzpicture}
\caption{Rewriting Fig.~\ref{left}}
\label{left1}
\end{center}
\end{figure}

\begin{figure}[H]
\begin{center}
\begin{tikzpicture}
\draw [thick, ->] (7.5,2.5)--(6.5,2.5);
\draw [thick, ->] (8,1)--(6,1);
\draw [thick, ->] (2.5,2.5)--(3.5,2.5);
\draw [thick, ->] (2,1)--(4,1);
\draw [thick, ->] (1,2)--(1,1.5);
\draw [thick, ->] (5,2)--(5,1.5);
\draw [thick, ->] (9,2)--(9,1.5);
\draw (3,2.5)node[above]{$S_4$};
\draw (7,2.5)node[above]{$T$};
\draw (9,1.75)node[right]{$S_7$};
\draw (5,1.75)node[right]{$S_9$};
\draw (1,1.75)node[left]{$S_2$};
\draw (3,1)node[below]{$S_6$};
\draw (7,1)node[below]{$S_8$};
\draw (1,2.5)node{$C\otimes X_1\otimes D$};
\draw (5,2.5)node{$C\otimes X_2$};
\draw (9,2.5)node{$B\otimes A \otimes X_2$};
\draw (1,1)node{$X_5 \otimes D$};
\draw (5,1)node{$X_6$};
\draw (9,1)node{$B\otimes X_4$};
\end{tikzpicture}
\caption{The right hand side}
\label{right}
\end{center}
\end{figure}

Conversely, suppose that
a pair $(F, F')$ of 3-tensors satisfies the half zipper
condition for $a_A, a_B, a_C$.
Equivalence of (1) and (4) in \cite[Theorem 4.1]{K6}
implies that the pair $(F, F')$ gives a flat field of strings
for a bi-unitary connection $W_B\otimes W_A\oplus W_C$.
The proof of \cite[Theorem 3.3]{S} describes flat fields of
strings arising from this type of bi-unitary connections, 
and it shows that the flat fields of strings are 
in the image of 
$T\in\Hom(B\otimes A,C)$ under the map 
\begin{align*}
&T\in\Hom((B\otimes A)\oplus C, (B\otimes A)\oplus C)\\
\mapsto \bigoplus_X\; &T\otimes \id_X\in\bigoplus_X
\Hom(((B\otimes A)\oplus C)\otimes X, ((B\otimes A)\oplus C)\otimes X),
\end{align*}
where $X$ runs through equivalence classes of irreducible
$M$-$P$ bimodules in the unitary multi-fusion category.
We thus have a corresponding coisometric intertwiner
$T\in\Hom(B\otimes A,C)$.  For an irreducible
$M$-$Q$ bimodule $X_2$,
$N$-$Q$ bimodule $X_4$, and
$L$-$Q$ bimodule $X_6$
in the unitary multi-fusion category,
we choose coisometric intertwiners
$S_7\in\Hom(B\otimes A\otimes X_2, B\otimes X_4)$,
$S_8\in\Hom(B\otimes X_4, X_6)$,
$S_9\in\Hom(C\otimes X_2, X_6)$.
We expand $T\otimes\id_{X_2}$ in terms of a
basis vector $(S_8\cdot S_7,S_9)$ in the string algebra,
then obtain the coefficient given by $W'_{T}$ in Fig.~\ref{q6T}.
We similarly obtain $W_{T}$ in Fig.~\ref{q6T} and the two
coefficients satisfy the identity of half flatness as in
\cite[Fig.~33]{K6}.  (Here $T$ needs not to be coisometric.)
We finally conclude that our pair $(F,F')$ is
of the form $(F_T, F'_T)$ arising from this $T$ 
in the above way because of the convention in Fig.~\ref{threeF}.
\end{proof}

\begin{remark}{\rm
We dealt with the half zipper condition rather than
the zipper condition in the above Theorem, but we
can easily translate this into the framework of
the zipper condition (and flatness rather than
half flatness) as in \cite[Theorem 4.1]{K6}.
}\end{remark}

\section*{Acknowledgements}

This work is supported by Grants-in-Aid 
for Scientific Research 24K21514 and 25H00590, and
RIKEN Pioneering Project
``Mathematical foundation of quantum information''.
A part of this work was done at
Fordham University and Perimeter Institute, and the
author is grateful for their hospitality.
The author also acknowledges supports of
U.S. Army Research Office through 
contract W911NF-25-1-0050.

\section*{Data Availability}

All data generated or analyzed during this study are 
included in this article.

\section*{Conflict of interest} 

The author declares that there is no conflict of 
interest regarding the publication of this manuscript.

\section*{The AI disclosure} 

The author completed this manuscript without using AI first,
asked ChatGPT 6.1 Sol for mathematical and English language
problems, and fixed minor issues pointed out.

\end{document}